\documentclass[12pt]{article}
\usepackage{cite}
\usepackage{amsmath,amsthm, amscd, amssymb, amsfonts}
\usepackage{appendix}
\usepackage{graphicx}
\usepackage[round]{natbib}
\begin{document}
\title{Comment on ``A Loophole of All ``Loophole-Free'' Bell-Type Theorems''}
\author{Justo Pastor Lambare}
\date{}
\maketitle
\begin{abstract}
In a recent article(Found Sci (2020) https://doi.org/10.1007/s10699-020-09666-0) Marek Czachor claims that the  Bell inequality cannot be proved because variables of complementary measurements cannot be added or multiplied. Even though he has correctly identified the problems existing with the orthodox interpretation of the Bell inequality and dealt with them in an original way, the interpretation he addresses do not pertain to the original formulation given by John Stewart Bell.
\end{abstract}
%

%
\section{Introduction}\label{sec:intro}
We shall refer to the Clauser, Horne, Shimony, and Hold(\citeyear{pCHSH69}) inequality  indistinctively as the CHSH inequality or Bell inequality.

We explain why \cite{pCza20a} views are not applicable in the context the Bell inequality was conceived, however, the discussion has a general relevance because Czachor's interpretation of the CHSH inequality is shared by most physicists although the majority of them, unlike Czachor, do not seem to be aware of the inapplicability of such an interpretation.
\section{The Orthodox Interpretation}\label{sec:toi}
Marek Czachor has ``hit the nail on the head'' when criticizing the usual interpretation asserting:
\begin{quote}
\emph{One can prove a counterfactual inequality, but it may not apply to actually performed measurements}.
\end{quote}
Although this remark encapsulates the core of the problem with the orthodox interpretation of the Bell inequality, it belongs  neither to any of the formulations made by John Bell nor to the formulation of \cite{pCHSH69}. Perhaps a more suitable title for Czachor's paper would be ``A Loophole of All ``Counterfactual''  Bell-Type Theorems''.

To our knowledge, the first to propose the inclusion of counterfactuals was \cite{pSta71} and it was soon accepted by most physicists, maybe because it simplified even further the already simple derivation of the Bell inequality.

The introduction of \emph{counterfactual definiteness} into the Bell theorem has generated a persisting debate about the problems and limitations brought by the use of subjunctive conditionals when dealing with physically objective situations. Contrary to what is commonly believed, the problems introduced by this kind of speculations are unrelated to quantum mechanics or classical physics, they are related to objective logical and mathematical inconsistencies, some of which, Czachor correctly points out.
\section{The Orthodox Mistake}\label{sec:tom}
 Derivations based on counterfactual results begin by writing the actual outcomes, say $A_1^{(j)}$, $B_1^{(j)}$ registered by Alice and Bob in the $j^{\,\underline{th}}$ run  of the experiment, then adding three counterfactual terms
\begin{eqnarray}
s^{(j)}  &=& A_1^{(j)}B_1^{(j)} -\underbrace{A_1^{(j)}B_2^{(j)}+A_2^{(j)}B_1^{(j)}+A_2^{(j)}B_2^{(j)}}_{counterfactual\quad results}\label{eq:act1}\\
\nonumber\\
         &=& A_1^{(j)}(B_1^{(j)} -B_2^{(j)})+A_2^{(j)}(B_1^{(j)} + B_2^{(j)})\label{eq:act2}
\end{eqnarray}
the result is $s^{(j)}=\pm2$ whatever the initial actual value $A_1^{(j)}B_1^{(j)}$. For $N$ runs $j=1,2\ldots,N$ the average is
\begin{equation}\label{eq:fr}
\begin{array}{c}
-2\leq\frac{1}{N}\sum_j s^{(j)} \leq 2\\
\\
-2\leq E(A_1B_1) - E(A_1B_2) + E(A_1B_2) + E(A_2B_2)\leq 2
\end{array}
\end{equation}
Where
\begin{equation}
E(A_iB_k)=\lim_{N\rightarrow\infty}\frac{1}{N}\sum_{j=1}^{N} A_i^{(j)}B_k^{(j)}\,;\quad i,k\in\{1,2\}
\end{equation}
The key problem with the dubious methodology followed in this derivation, and which was correctly spotted by Czachor, concerns the testability of the predicted results or, in other words, how are the counterfactual results supposed to be replicated by actual experiments?
The three counterfactual terms in (\ref{eq:act1}) are spurious and their inclusion modifies the result corresponding to the values obtained in the real experiment.

Of course, one can add hypothetical terms to a mathematical expression, however, in those cases it is necessary that those terms add to zero. This does not happen in (\ref{eq:act1}) because three terms with values $\pm1$ cannot cancel each other.

The problem, as Czachor correctly points out, is purely logical in nature and has nothing to do with classical physics or quantum mechanics.
\section{Czachor's Proposal}\label{sec:cp}
M. Czachor proposes to trace back the problems with the usual interpretation of the Bell inequality to non-Diophantine arithmetics which would be analogous to the arithmetic of eigenvalues of non-commuting observables. However, the formal example in section 2 of \cite{pCza20a} may be criticized by followers of the orthodox view in that it does not comply with one of the Bell theorem hypotheses, namely, \emph{measurement independence}.

\emph{Measurement independence} is the assumption that the probability distribution function $\rho$ is independent from the device settings, and is related to the \emph{free will} of the experimenters. Indeed, \emph{measurement independence}  implies that the domain of the probability function cannot change when the settings are changed.
In Bell's probabilistic model, the correlation term is given by
\begin{eqnarray}
E(A_iB_k)=\int \rho(\lambda)A_i(\lambda)B_k(\lambda)\,d\lambda\label{eq:bct}
\end{eqnarray}
Notice that the distribution function in (\ref{eq:bct}) is written as $\rho(\lambda)$, not $\rho_{ik}(\lambda)$. This implies the domain of $\rho$ must remain the same when the settings $i,k$ are changed. Equations (five) and (six) of Czachor's model violate this condition; the domain of $\rho_{00}$ is the rectangle $(0,1)\times (0,1)$ while the domain of $\rho_{10}$ is $(1,2)\times (0,1)$.

Thus, it seems likely that advocates of the orthodox views would dismiss Czachor's justified criticisms as another unfounded misinterpretation.

It is necessary to point out that the Bell theorem does not forbid the existence of local realistic models that violate the Bell inequality. In fact, the Bell theorem states the inequality cannot be violated by a local realistic model that satisfies all the hypothesis of the theorem. One of those hypotheses is \emph{measurement independence}. Thus, although Czachor has successfully constructed an explicit model \citep{pCza20b} that violates the Bell inequality, such a model cannot be considered to be a disproof of the Bell theorem.

\cite{pCza20b} has devised a mathematically interesting local realistic model that violates the Bell inequality but does not violate the Bell theorem. Whether such a model is physically meaningful constitutes a different subject that perhaps deserves a different type of discussion. 
\section{The Correct Interpretation}\label{sec:tci}
The correct derivation does not start by writing individual results as in (\ref{eq:act1}).
Correct derivations start by writing the expectation values of actual results, as can be seen in any of the writings of John Stewart Bell or in \cite{pCHSH69}, and shown ahead in equations (\ref{eq:s1}) through (\ref{eq:s6}).

An analysis of such derivations traces back the attainment of the the bound value $2$ of the inequality, not to the use of counterfactual results, but to the hypothesis of \emph{measurement independence}. Although \emph{measurement independence} is commonly related to the \emph{free will assumption}, it also implies a statistical regularity that \cite{pDBae84} termed the \emph{reproducibility hypothesis}
\footnote{Puzzled by many derivations that start by (\ref{eq:act1}) or (\ref{eq:cs}), presenting the Bell inequality as a trivial ``Fact''\citep{pGil14}, \cite{pLam17} explains the correct interpretation being unaware of concepts such as \emph{measurement independence}, the \emph{reproducibility hypothesis} and \emph{counterfactual definiteness}.}.

To see how the \emph{reproducibility hypothesis} is related to \emph{measurement independence} we reproduce the steps of a standard derivation of the inequality
\begin{eqnarray}
S    &=& E(a_1,b_1)-E(a_1,b_2)+E(a_2,b_1)+E(a_2,b_2)\label{eq:s1}\\
     &=& \int \rho(\lambda)\,C(\lambda)\,d\lambda\label{eq:s2}\\
|S|  &\leq& \int \rho(\lambda)\,|C(\lambda)|\,d\lambda\label{eq:s3}\\
     &\leq& \int \rho(\lambda)\, 2\,d\lambda\label{eq:s4}\\
     &\leq& 2\int \rho(\lambda)\,d\lambda\label{eq:s5}\\
     &\leq& 2\label{eq:s6}
\end{eqnarray}
The term $C(\lambda)$ in (\ref{eq:s2}) is given by
{\small
\begin{eqnarray}\label{eq:cs} 
  C(\lambda)=A(a_1,\lambda)B(b_1,\lambda)-A(a_1,\lambda)B(b_2,\lambda)+A(a_2,\lambda)B(b_1,\lambda)+A(a_2,\lambda)B(b_2,\lambda)
\end{eqnarray}
}
The last expression appears as a mathematical consequence of taking $\rho$ as a common factor, and it would not be possible if $\rho$ contained the settings as explicit variables.

When (\ref{eq:cs}) is literally interpreted it seems to imply the need of four consecutive experiments with the same $\lambda$. Since the experimenter has no control over the hidden variables, it is impossible to actually reproduce such experiments\footnote{In fact, we do not even know if such hidden variables exist, it is the result found in the experiment that falsifies their existence.}. This impossibility has lead to the following interpretation\citep{pSta71}: \emph{``Of these eight numbers only two can be compared directly to experiment. The other six correspond to the three alternative experiments that could have been performed but were not''}.

The previous interpretation, based on \emph{counterfactual definiteness}, can be avoided by realizing that the experiment does not need to reproduce (\ref{eq:cs}) in a direct way.

Really, to reproduce (\ref{eq:s1}), we only need to perform individual experiments where joint actual measurements produce one of the four possible outcomes $A_1B_1$, $A_1B_2$, $A_2B_1$, $A_2B_2$. After a long series of trials we can collect sufficient data to evaluate the expectation values $E(A_iB_k)$.

If the assumption of measurement independence is correct and if the local realist functions $A_i(\lambda)$, $B_k(\lambda)$ in (\ref{eq:bct}) actually exist, we can expect the values of $\lambda$ to be regularly repeated in different experiments with different settings, allowing the appearance of (\ref{eq:cs}) after conveniently rearranging the actual data according to the rules of arithmetics.

This is the explicit physical interpretation of the mathematical operations that allow the passage from (\ref{eq:s1}) to (\ref{eq:s2}), resulting in the emergence of (\ref{eq:cs}). We term this interpretation, along with Willy De Baere, the \emph{reproducibility hypothesis}\footnote{We do not know if De Baere would agree with this interpretation, but we do credit him for the idea and the terminology.}.

Of course, the rearrangement of data in (\ref{eq:s1}) to obtain (\ref{eq:s2}), is purely conceptual but it is falsified by the result of the experiment.

The long-disregarded \emph{reproducibility hypothesis} turns out to be of crucial importance to avoid inconsistencies and make sense of Bell's derivation without invoking the unjustified materialization of counterfactual experiments, as correctly criticized by Czachor and mathematically proved proved by \cite{pLam19}\footnote{This manuscript contains two theorems proving the physical irrelevance of the \emph{counterfactual definiteness} hypothesis.}.
\section{Measurement Independence and Free Will}\label{sec:miafw}
In the 1964 original version of the Bell theorem\citep{pBel64}, \emph{measurement independence} was an implicit assumption; later in 1975 John Bell(\citeyear{pBel76}) recognized \emph{measurement independence} as a necessary independent assumption and justified it by the freedom of the experimenters for choosing their settings.

He reasoned that if $p(ab)$ is the joint probability of Alice and Bob for choosing settings $a$ and $b$, then freedom requires that
    \begin{equation}\label{eq:fw} 
          p(ab|_\lambda) = p(ab)
    \end{equation}
According to Bell \citep{pBel81}:
\begin{quote}
``..., we cannot be sure that $a$ and $b$ are not significantly influenced by the same factors $\lambda$ that influence $A$ and $B$. But this way of arranging quantum mechanical correlations would be even more mind boggling than one in which casual chains go faster than light.''
\end{quote}
According to Bayes's theorem of probability theory
\begin{equation}\label{eq:bayes}
p(ab\lambda)=p(ab|_\lambda)\rho(\lambda)=\rho(\lambda|_{ab})p(ab)
\end{equation}
(\ref{eq:fw}) and (\ref{eq:bayes}) give \emph{measurement independence}
    \begin{equation}\label{eq:mi} 
        \rho_{ab}(\lambda)\equiv\rho(\lambda|_{ab})=\rho(\lambda)
    \end{equation}
Hence, if we reject \emph{measurement independence}, we must also jettison freedom, a stance which is known as superdeterminism.
\section{Czachor's Formal Example and Free Will}\label{sec:feafw}
Although Czachor's example\citep{pCza20a} violates \emph{measurement independence}, it is worth noticing that it cannot be deemed logically incorrect if superdeterminism is explicitly recognized.

However, Czachor claims that his model respects freedom so, if correct, it would be a proof by counterexample that the reasoning that leads from (\ref{eq:fw}) to (\ref{eq:mi}) is flawed. We shall see, nonetheless, that a close examination of Czachor's nice example reveals that it violates freedom if the rules of the ``Bell game'' are applied correctly and without cheating.

In the case under analysis $\lambda=(x,y)$, where the hidden variable associated to Alice is $x$ and the one associated to Bob is $y$. The settings are determined by the values $\alpha=0$ and $\alpha=1$ giving $A_0\,,\,A_1$ and $B_0\,,\,B_1$ for Alice and Bob respectively. It is necessary to notice that, although Alice and Bob should be able to choose freely between their two options, those options $A_0, A_1$ and $B_0, B_1$ are fixed in advanced and cannot change during the experiment.

The local realistic functions $a_\alpha(\lambda)$, $b_\beta(\lambda)$ that determine the results are given by equations (two) and (three) of \cite{pCza20a}.

With respect to the meaning of the hidden variables, already in \cite{pBel64} we find:
\begin{quote}
``In a complete physical theory of the type envisaged by Einstein, the hidden variables would have dynamical significance and laws of motion; our $\lambda$ can be thought of as initial values of these variables at some suitable instant.''
\end{quote}
So, if $\lambda'(t)$ are these dynamical variables, the hidden variables of the Bell theorem would be their initial values $\lambda=\lambda'(t_0)$ where the ``suitable instant''  $t=t_0$ could be the moment the pair is created or the moment the particles leave the source. The important point is that, in Bell's local realistic model, those variables are fixed previously to the measurements as events belonging to their common past light cones and they cannot change after they were fixed.

In our case, this means that each time Alice leaves her home, she carries the instruction to stay at room number $x$, i. e., the value of $x$ is fixed before she makes her local decision between $\alpha=0$ or $\alpha=1$.

Later, when she arrives at Acetown the problem with her freedom becomes evident. According to the model, $0< x< 2$ then if her hidden variable is $x<1$, since we have the restriction $\alpha<x<\alpha+1$, she cannot freely choose between $\alpha=0$ and $\alpha=1$ because $\alpha=1$ is forbidden to her. Analogously, if $x>1$, she cannot choose $\alpha=0$, and when $x=1$, she cannot choose either hotel!

Thus, we see that in this particular case, violation of measurement independence severely compromises the freedom of the experimenters.

Of course, if we let Alice fix the value of her room($x$) after she chose her hotel($\alpha$), she would regain her free will; however, we would be breaking the rules of the ``Bell game''.

We can complain about the rules of the game, but those rules were not capriciously established. To understand the reasons behind those rules, we must go back to the real physical problem and remember that they were set with the purpose of finding a local explanation for the puzzling existence of perfect correlations. Thus, we are forced to see the hidden variables as the common causes lying in the common casual past of the measuring events.
\section{Conclusions}
The Bell theorem is an elementary mathematical result based on three assumptions, namely, locality, realism, and measurement independence; thus, respecting those assumptions under a correct interpretation, it cannot present any loophole.

One possible incorrect interpretation, as Czachor has pointed out, is \emph{counterfactual definiteness}.
Another popular logically incorrect view confounds conditions for the existence of joint probabilities as excluding the nonlocality implications of the Bell inequalities violations\footnote{Sometimes this view takes the form of a presumed hidden assumption made by Bell, for a discussion see \cite{pLam17b}.}.

However, apart from faulty interpretations, logically correct rejections of the Bell theorem are possible and they are based on the exclusion of measurement independence as an appropriate physical assumption.

Although Czachor has conflated some inconsistent views with others that can be considered logically correct, by separating them, we hope to have clarified some important points.

Particularly important is that Marek Czachor has correctly identified the inconsistent nature of the orthodox interpretation of the Bell inequality, however, it is necessary to vindicate John Stewart Bell against such ``counterfactual'' interpretations that do not belong to the correct formulation of the inequality as originally developed by \cite{pBel64} and \cite{pCHSH69}.
\section*{Conflict of Interest}
On behalf of all authors, the corresponding author states that there is no conflict of interest.
%

%
\end{document}